\documentclass{article}
\usepackage{ijcai23}

\usepackage{times}
\usepackage{soul}
\usepackage{url}
\usepackage[hidelinks]{hyperref}
\usepackage[utf8]{inputenc}
\usepackage[small]{caption}
\usepackage{graphicx}
\graphicspath{{results/figs/}}
\usepackage{amsmath}
\usepackage{amsthm}
\usepackage{booktabs}
\usepackage{multirow}
\usepackage{tabularx}
\include{dcolumn}
\usepackage[switch]{lineno}

\usepackage{subfiles}
\usepackage{amssymb}
\usepackage{thmtools}
\usepackage{thm-restate}

\usepackage{balance}
\usepackage{import}
\usepackage{calrsfs}
\DeclareMathAlphabet{\pazocal}{OMS}{zplm}{m}{n}

\usepackage{tikz,pgffor}
\usetikzlibrary{arrows,backgrounds,calc}
\usetikzlibrary{automata,positioning,arrows,shapes,math,arrows.meta,decorations.pathmorphing}
\tikzstyle{min}=[thick,circle,draw,minimum size=1.3em,inner sep=0em,text centered]

\newcommand{\tm}{\mathit{time}}
\newcommand{\Nset}{\mathbb{N}}

\newcommand{\Rset}{\mathbb{R}}

\newcommand{\Inv}{\mathbb{I}}

\newcommand{\MP}{\mathrm{MP}}
\newcommand{\Val}{\mathrm{Val}}
\newcommand{\Mem}{M}
\newcommand{\prob}{\mathbb{P}}

\newcommand{\ag}[1]{\widehat{#1}}
\newcommand{\size}[1]{|\!|{#1}|\!|}

\newcommand{\calD}{\mathcal{D}}

\newcommand{\calO}{\mathcal{O}}
\newcommand{\calS}{\mathcal{S}}

\newcommand{\PSPACE}{$\mathsf{PSPACE}$}
\newcommand{\NP}{$\mathsf{NP}$}

\usepackage{xspace}
\makeatletter
\DeclareRobustCommand\onedot{\futurelet\@let@token\@onedot}
\def\@onedot{\ifx\@let@token.\else.\null\fi\xspace}
\def\eg{e.g\onedot}
\def\ie{i.e\onedot}
\def\cf{cf\onedot}

\makeatother

\definecolor{ourblue}{RGB}{53, 95, 141}
\hypersetup{
	final,
	colorlinks,
	breaklinks=true,
	linkcolor=ourblue,
	citecolor=ourblue,
	urlcolor=ourblue,
}

\usepackage{algorithm}
\usepackage[noend]{algpseudocode}

\newtheorem{remark}{Remark}
\newtheorem{proposition}{Proposition}

\title{Mean Payoff Optimization for Systems of Periodic Service and Maintenance}

\author{
David Kla\v{s}ka\and
Anton\'{\i}n Ku\v{c}era\and 
V\'{\i}t Musil\and  
Vojt\v{e}ch \v{R}eh\'{a}k
\affiliations
Masaryk University, Brno, Czech Republic
\emails
tony@fi.muni.cz
}

\begin{document}

\maketitle

\begin{abstract}
Consider oriented graph nodes requiring periodic visits by a service agent.
The agent moves among the nodes and receives a payoff for each completed service task, depending on the time elapsed since the previous visit to a node.
We consider the problem of finding a suitable schedule for the agent to maximize its long-run average payoff per time unit.
We show that the problem of constructing an \mbox{$\varepsilon$-optimal} schedule is \PSPACE-hard for every fixed $\varepsilon \geq 0$, and that there exists an optimal \emph{periodic} schedule of exponential length.
We propose \emph{randomized finite-memory (RFM)} schedules as a compact description of the agent's strategies and design an efficient algorithm for constructing RFM schedules.
Furthermore, we construct deterministic periodic schedules by sampling from RFM schedules.
\end{abstract}

\section{Introduction}
\label{sec-intro}

Workforce scheduling and routing problems (WSRP) \cite{CSLSQ:workforce-sched-survey-AOR} refer to scenarios involving mobile personnel performing periodic service or maintenance at different locations, where the transport time significantly influences the overall efficiency.
Hence, WSRP is a combination of employee scheduling and vehicle routing problems (VRP) \cite{TV:book}, where the latter provides the methodology for computing appropriate trajectories for service agents.
Since services can be performed at customers' locations only in certain time intervals, the \emph{vehicle routing problem with time windows (VRPTW)} \cite{KLMS:vehicle-rout-windows} is particularly relevant in this setting.

An instance of VRPTW is a finite set of nodes $S$ where each $v \in S$ is assigned a time interval $[a_v,b_v]$.
A vehicle must arrive to~$v$ before time~$b_v$, and if it arrives before time $a_v$, it must wait until~$a_v$.
Moving between $v,u$ takes time $\tm(v,u)$, and costs $c(v,u)$.
The task is to design a set of routes for a given number of vehicles minimizing the total costs such that each node is visited precisely once and on time.
The routes are obliged to start/end at designated nodes (depots), and there is a time horizon $b$ bounding the length of every route such that $b_v \leq b$ for every~$v$.
Intuitively, the $b$ corresponds to one working shift.

\paragraph{Motivation.}
Although VRPTW is appropriate for modeling the ``routing component'' of many periodic maintenance problems, it is not applicable in situations when the maintenance period demanded by a customer is significantly larger than one working shift, and the preference for the period is specified by a non-trivial payoff function rather than just lower/upper bounds.
To understand the need to overcome these limitations, consider the following simple scenario.

Let $S$ be a set of public vending machines requiring periodic maintenance.
For each machine $v$, let $r_v$ be the time after which $v$ needs another preventive service action.
Note that $r_v$ may range from days to weeks.
The owner of the machines demands a maintenance period $r_v$ for every~$v$ and is willing to pay $p_v$ for each service action, \ie spend $\frac{p_v}{r_v}$ per time unit for maintenance costs.
If~$v$ is serviced prematurely at time $t < r_v$, the owner has no reason to complain, but it insists on keeping the same maintenance costs per time unit, i.e., the agent receives only $t \cdot \frac{p_v}{r_v}$ for this service.
If~$v$ is serviced at time $t > r_v$, the owner charges a negative penalty $c_v < 0$ for every time unit exceeding $r_v$, i.e., the agent receives $p_v + (t-r_v)\cdot c_v$.
The owner may also introduce extra penalties for very short times.
The aim of the agent is to \emph{maximize the average payoff per time unit}.
To avoid costly premature returns to certain nodes, the agent may prolong the actual traversal time between two nodes by non-negative \emph{waits}.

Note that the above scenario \emph{does not impose any bounded planning horizon}.
The agent may still be required to return to a depot after $b$ time units (a working shift), which can be modelled by a node where significant deviations from $b$ are ``punished'' by a negative payoff.
Also, note that the payoff function can be even more complicated in scenarios where the timing constraints are influenced by multiple criteria.

\paragraph{Our contribution.}
We introduce and study the \emph{infinite horizon recurrent routing problem (IHRRP)} specified as a tuple $\calS = (S,\tm,\{P_v : v \in S\},D)$ where $S$ is a finite set of nodes, $\tm(v,u)$ is the traversal time between the nodes $v$ and $u$, $P_v(t)$ is a \emph{payoff} received by a service agent when $v$ is revisited after exactly $t$ time units, and $D \subseteq S \times S$ is a subset of moves along which the agent is allowed to wait.
The problem is to compute an appropriate \emph{schedule} (moving plan) for the agent maximizing the \emph{mean payoff}, \ie the long-run average payoff per time unit.
Our main results (1)--(4) are summarized below.
\begin{itemize}
\item[(1)] We show (Theorem~\ref{thm-periodic}) that for every $\varepsilon \geq 0$, there exists an \emph{$\varepsilon$-optimal} schedule which is \emph{deterministic} and \emph{periodic}. Such a schedule can be represented as a finite cycle over the nodes, and the length of this cycle is at most \emph{singly exponential} in size of~$\calS$.
\end{itemize}
\begin{itemize}
\item[(2)] We prove (Theorem~\ref{thm-PSPACE-hard}) that solving IHRRP is \PSPACE-hard, even when demanding only sub-optimal solutions. More precisely, it is \PSPACE-hard to distinguish whether the best mean payoff achievable for a given instance $\calS$ is at least~$1$ or at most $1-\varepsilon$ for arbitrary $\varepsilon \geq 0$.
\end{itemize}
According to~(2), the optimal achievable mean payoff is not only hard to compute but also hard to \emph{approximate} up to an arbitrary fixed $\varepsilon \geq 0$.
Hence, there is no efficient algorithm for constructing an $\varepsilon$-optimal schedule (randomized or deterministic), assuming $\mathsf{P} \neq \mathsf{PSPACE}$.
Furthermore, (2)~implies that the length of a cycle representing an \mbox{$\varepsilon$-optimal} deterministic and periodic schedule of~(1) \emph{cannot be bounded by any polynomial}, unless $\mathsf{NP} = \mathsf{PSPACE}$.
Hence, the exponential upper bound on its length established in~(1) is matched by the superpolynomial lower bound.

In principle, the upper bound of (1)~allows for constructing an \mbox{$\varepsilon$-optimal} deterministic schedule by the methods invented for finite-horizon routing problems.
These methods are mostly based on solving appropriate mathematical programs whose size is proportional to the horizon, and a solution is computable in \NP.
Since we need to consider an exponentially large horizon for IHRRP, the size of these programs becomes exponential, making the total running time even doubly exponential.
This explains why the techniques for finite-horizon routing problems are not applicable to the infinite-horizon case.
In this work, we use a different algorithmic approach.

\begin{itemize}
\item[(3)] We introduce the concept of \emph{randomized finite-memory (RFM) schedules} and design a polynomial-time algorithm for constructing a RFM schedule for a given IHRRP instance based on differentiable programming and gradient ascent (Algorithm~\ref{alg:optim}).
\end{itemize}
Finite-memory strategies are equipped with a finite set $M$ of memory states.
In each step, the agent's (possibly randomized) decision depends on its current location
$v\in S$ and its current memory state $m\in M$.
Intuitively, the memory states are used to ``remember'' some finite information about the sequence of previous moves executed by the agent.
In general, an \mbox{$\varepsilon$-optimal} RFM schedule may require memory with exponentially many states. However, randomization helps to reduce this blowup.
RFM schedules can achieve a substantially higher mean payoff than deterministic schedules with the same (or even larger) memory.
This is the crucial tool for tackling the \PSPACE-hardness of the IHRRP problem. 

Randomized schedules are easily executable by robotic devices, but they are less appropriate for human agents\footnote{Previous works on finite horizon routing problems consider only deterministic strategies.}.
Therefore, we also consider the problem of constructing a \emph{deterministic} schedule.

\begin{itemize}
\item[(4)] We design an algorithm for constructing a deterministic periodic schedule for a given IHRRP instance based on identifying optimal cycles in long routes generated by a previously computed RFM schedule (Algorithm~\ref{alg:determinize}). 
\end{itemize}
The routes are obtained by repeatedly ``executing'' a constructed RFM schedule. As we shall see in Section~\ref{sec-RFM-schedules}, RFM schedules are \emph{ergodic}, and the probability that a route of length $n$ contains a cycle achieving the same or even better mean payoff as the considered RFM schedule converges to~one as $n$ approaches infinity.

The algorithms of~(3) and~(4) are evaluated on instances of increasing size.
We use planar grids with randomly positioned nodes to avoid bias towards simple instances.
The algorithms process relatively large instances and produce high-quality schedules.
The quality of cycles discovered by the algorithm of~(4) grows with the quality of RFM schedules used to generate long routes.
The length of the resulting cycles exceeds $12$ working shifts, and the corresponding mean payoff is not achievable by short cycles corresponding to one shift (a typical planning horizon in finite-horizon routing problems).   

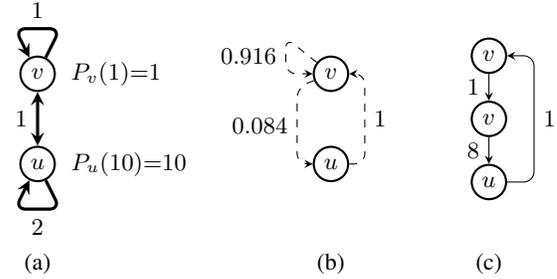
\begin{figure}[t]\centering
\begin{tikzpicture}[x=3cm, y=1.2cm,font=\footnotesize]
\foreach \x/\c/\l in {0/0/a,1/1.3/b,2/2/c}{%
    \coordinate (a) at (\c,0);
    \node at ($(a) +(0,-1.1)$) {(\l)};

    \ifthenelse{\x=2}{\node [min] (A) at ($ (a) + (0,1.2) $) {$v$};
                      \node [min] (B) at ($ (a) + (0,.5) $) {$v$};
                      \node [min] (C) at ($ (a) + (0,-.2) $) {$u$};
                      \draw[-stealth] (A) -- node[left] {$1$}  (B);
                      \draw[-stealth] (B) -- node[left] {$8$}  (C); 
                      \draw [-stealth,rounded corners] (C) -- +(.2,0) --node[right] {$1$} +(.2,1.4) -- (A);
                     }{
                      \node [min] (T1\x) at (a) {$u$};
                      \node [min] (T2\x) at ($ (a) + (0,1) $) {$v$};   
                     }
    \ifthenelse{\x=0}{\node [right=.5ex of T1\x] {\small$P_u(10){=}10$};
                      \node [right=.5ex of T2\x] {\small$P_v(1) {=}1$};
                      \draw[stealth-stealth,very thick] (T1\x) -- node[left] {$1$}  (T2\x);
                      \draw [-stealth,very thick,rounded corners] (T1\x) -- +(.1,-.5) --node[below] {$2$} +(-.1,-.5) -- (T1\x);  
                      \draw [-stealth,very thick,rounded corners] (T2\x) -- +(.1,.5) --node[above] {$1$} +(-.1,.5) -- (T2\x);  
                     }{}
    \ifthenelse{\x=1}{% 
    \draw[-stealth,rounded corners,dashed] (T1\x) -- ($(T1\x) +(.15,0)$) -- node[right]{$1$} ($(T2\x)+(.15,0)$) -- (T2\x); 
    \draw[-stealth,rounded corners,dashed] (T2\x) -- ($(T2\x) +(-.15,-.15)$) -- node[left]{$0.084$} ($(T1\x) +(-.15,0)$) -- (T1\x);
    \draw [-stealth,rounded corners,dashed] (T2\x) -- +(-.2, .4) -- node[left]{$0.916$} +(-.2,0) -- (T2\x);}{} 
}
\end{tikzpicture}
\caption{The deterministic and periodic schedule~(c) for the scenario~(a) requires $|M|=2$ and its mean payoff is $1.1$. 
A randomized schedule~(b) with $|M|=1$ yields mean payoff $1.307$, and the optimal periodic schedule (requiring $|M|=9$) can be discovered by random sampling.} 
\label{fig-rand}
\end{figure}

\paragraph{Example.}
To get some intuition about our results, consider an instance $\calS$  of Fig.~\ref{fig-rand}~(a) with two nodes $v$, $u$.
The traversal times are indicated by transition labels.
The payoff functions $P_v$, $P_u$ satisfy $P_v(t) = 1$ for all $t \geq 1$, and $P_u(t)$ is either $10$ or $0$ depending on whether $t \geq 10$ or not, respectively.
The agent can wait along all transitions.
\medskip

A simple RFM schedule is shown in Fig~\ref{fig-rand}~(b).
Since this schedule has $|M|=1$, it corresponds to a \emph{positional} (\emph{Markovian}) strategy.
For example, whenever the agent visits~$v$, it either performs a self-loop on $v$ or moves to $u$ with probability $0.916$ and $0.084$, respectively.
This trivial RFM schedule achieves a mean payoff~$1.307$.
The best \emph{deterministic} schedule with $|M|=1$ performs the self-loop on~$v$ forever, and the associated mean payoff is equal to~$1$.
A deterministic schedule with $|M|=2$ can achieve a mean payoff equal to $1.1$ in the way shown in Fig~\ref{fig-rand}~(c).
The schedule performs one self-loop on $v$, then moves to $u$ (prolonging the traversal time to $8$ by waiting), and then returns back to $v$.
This cycle is repeated forever.
Note that two memory states are needed to ``remember'' whether the self-loop on $v$ has already been performed or not.

In fact, a deterministic schedule needs $|M|\geq 4$ to achieve a mean payoff better than~$1.307$.
Hence, even in this trivial example, \emph{using randomization with $|M|=1$ achieves better mean payoff than deterministic schedules with $|M|=3$}, illustrating the crucial advantage of RFM schedules mentioned above.

Now, let us consider the algorithm of~(4) when the constructed RFM schedule with $|M|=1$ is used to generate long routes in the considered instance.
Due to the large probability of the self-loop on $v$, a randomly generated route initiated in $v$ tends to start with ``many'' copies of $v$ eventually followed by $u$.
Hence, the \emph{optimal} deterministic and periodic schedule with mean payoff $1.9$ represented by the cycle $v,1,v,1,v,1,v,1,v,1,v,1,v,1,v,1,v,1,u,1,v$ is discovered quickly.
This illustrates how the constructed RFM schedule ``navigates'' the search for a cycle with high mean payoff.

\goodbreak
\subsection{Related Work}
\label{sec-related}

Most of the existing works dealing with routing and maintenance optimization use the routing model to plan maintenance operations and thus divide the optimization task into two phases.
Examples include the Technician Routing and Scheduling Problem \cite{ZS:MTRSP-problem-EJOS}, the Technician and Task Scheduling Problem \cite{CLPR:sched-tech-telcompany-JS}, the Service Technician Routing and Scheduling Problem \cite{KPDH:adapt-neigh-tech-rout-JS}, or the Geographically Distributed Asset Maintenance Problem \cite{CHPCMR:risk-driven-maintenance}.
More recent works \cite{LS:combined-main-rout-RESS,JCHRK:integrated-rout-main-RESS} combine maintenance and routing models into one optimization framework.
We refer to \cite{CSLSQ:workforce-sched-survey-AOR,DKBGF:opt-main-rout} for more detailed overviews.

The above models are finite-horizon, and the existing solution approaches can be classified as (i) heuristics and meta-heuristics, (ii) mathematical programming, (iii) hybrid methods. Examples of heuristic methods are local search \cite{SVBO:team-orient-windows-TS}, adaptive large neighborhood search \cite{KPDH:adapt-neigh-tech-rout-JS,CLPR:sched-tech-telcompany-JS,PGM:heur-tech-rout-sched-OL}, tabu search~\cite{TMT:sched-ched-geog-TR}, greedy heuristics \cite{XCh:field-tech-sched-JH}, etc.
Mathematical programming approaches are mostly based on constructing a mixed integer program and then applying commercial solvers or using branch-and-price or similar algorithms (e.g., \cite{BDGT:multiperiod-planning,BFG:team-orient-4OR}).
Hybrid approaches are based, e.g., on the combination of linear programming, constraint programming, and metaheuristics \cite{BF:hybrid-health-COR}, or the combination of linear programming and repeating matching heuristics \cite{EFR:LapsCare-EJOR}.
A more comprehensive overview can be found in \cite{ZS:MTRSP-problem-EJOS}.   
 
Randomized strategies for infinite horizon path planning problems have been used in the context of adversarial patrolling~\cite{VAT:adversarial-patrolling,BGA:patrolling-arbitrary-topologies,KKMR:Regstar-UAI}.
Here, randomization is crucial for decreasing the predictability of the patroller.
Technically, the patrolling problem is a variant of recurrent bounded reachability optimization which is different from the considered IHRRP problem.

To the best of our knowledge, infinite horizon routing problems related to periodic maintenance have not yet been studied in previous works. 

\section{Mathematical Model}
\label{sec-model}

We use $\Nset$ and $\Nset_+$ to denote the sets of all non-negative and positive integers, respectively.
We assume familiarity with basic notions of probability theory and Markov chain theory (ergodicity, invariant distribution, etc.).

\subsection{Service Specification}
Let $S$ be a finite set of \emph{nodes}, and let $\tm\colon S \times S \rightarrow \Nset_+$ be a function such that $\tm(v,u)$ is the time needed to move from $v$ to $u$. 
Performing a service action at a node $v$ also takes some time, but we assume that this time has already been added to $\tm(v,u)$ for every $u$.
This also explains why $\tm(v,v)$ can be positive.

For every $v \in S$, let $P_v\colon\Nset_+ \to \Rset$ be a function such that $P_v(t)$ is the payoff received by a service agent when $v$ is revisited after time~$t$ units.
We require that $P_v$ is \emph{eventually affine}, i.e., there exist~$k_v \in \Nset_+$ and~$d_v,c_v \in \Rset$ such that $P_v(k_v{+}t) = d_v + t\cdot c_v$ for all $t \in \Nset$.
Note that no restrictions are imposed on the values of $P_v(t)$ for $t < k_v$.

The constant $k_v$ is typically chosen sufficiently large so that revisiting $v$ after more than $k_v$ time units is highly undesirable and $f(k_v{+}i)$ is negative for all $i \geq 0$.
Then, the agent is not motivated to re-visit $v$ after $k_v$ time units, and hence the precise value of $f(k_v{+}i)$ does not matter.
The exact purpose of the constant $c_v$ is clarified in Section~\ref{sec-schedules} (it is used to prevent the agent from suspending visits to a certain subset of nodes).

When an agent moves between $v$ and $u$, it may intentionally prolong the traversal time beyond $\tm(v,u)$ if this leads to a higher payoff.
Since the underlying transport infrastructure may allow for prolonging only certain moves, we specify a subset $D \subseteq S \times S$ of \emph{prolongable moves}.

A \emph{service specification} is defined as a tuple $\calS = (S,\tm,\{P_v : v \in S\},D)$.
Every $P_v$ is represented as a finite sequence $P_v(1),\ldots,P_v(k_v),d_v,c_v$, where all numbers are written in binary.
The \emph{enconding size of $\calS$}, denoted by~$\size{\calS}$, is the length of the (binary) string representing~$\calS$.
We also use $k_{\max}$ to denote $\max\{k_v : v \in S\}$.

\begin{remark}
Our model of service specification is \emph{intentionally simplified} so that all optimization criteria except for return times are eliminated, allowing for studying the ``timing aspects'' of the IHRRP problem in isolation. 
Additional features, such as \emph{costs} associated to moves, waits, service actions, etc., are straightforward extensions resulting in more technical (but not substantially different) variant of the objective function governing the algorithm presented in Section~\ref{sec-algo}. 
\end{remark}

\subsection{General Deterministic Schedules, Mean Payoff}
\label{sec-schedules}

A \emph{deterministic schedule} (or just a \emph{schedule}) is an infinite sequence $\alpha = v_1,\tau_1,v_2,\tau_2,v_3,\tau_3,\ldots$ where $v_i \in S$ and $\tau_i = \tm(v_i,v_{i+1}) + w_i$ is the time spent by moving from $v_i$ to $v_{i+1}$.
Here, $w_i \geq 0$ is an integer \emph{wait} prolonging the move.
We require that $w_i > 0$ only if $(v_i,v_{i+1}) \in D$.

Every schedule $\alpha$ determines the associated \emph{mean payoff} $\MP(\alpha)$, defined as the long-run average payoff per time unit.
For every $i\geq 1$, let $\ell_i$ be the index of the previous visit to $v_i$ (if there is no previous visit to $v_i$, then $\ell_i = 1$).
We also use $t_i$ to denote the total time elapsed since the previous visit to $v_i$, \ie $t_i = \sum_{j=\ell_i}^{i-1} \tau_j$.
Now we define
\begin{equation*}
    \MP(\alpha)  = \liminf_{n \to \infty} \frac{\sum_{i=1}^n P_{v_i}(t_i)}{\sum_{i=1}^n \tau_i} \,.
\end{equation*}
Clearly, $\MP(\alpha) = \MP(\alpha')$ for every infinite suffix $\alpha'$ of $\alpha$. 

In general, a schedule may avoid servicing some nodes and visit only a convenient subset of~$S$ generating a high mean payoff.
However, in some cases, the agent may be required to service a certain subset $C \subseteq S$ of \emph{compulsory nodes}  (such as the node modeling the depot, see Section~\ref{sec-intro}).
We implement this requirement as a \emph{soft constraint} in the sense that the agent may still avoid visiting $v \in C$ if it is willing to pay the penalty $c_v$ per every time unit.
If $c_v$ is set to a sufficiently negative value, then omitting $v$ results in a negative mean payoff.
Hence, the synthesis algorithm aims to avoid this situation by revisiting~$v$ infinitely often.
In some instances, setting the penalty $c_v$ to a ``moderately negative'' value makes sense.
For example, suppose the agent can hire an external provider for servicing~$v$.
In that case, the penalty $c_v$ may represent the long-run average costs generated by this subcontract, and suspending visits to a compulsory~$v$ may be an eligible rational decision for the agent.
For technical convenience, \emph{from now on we assume that $c_v = 0$ for all $v \not\in C$}.

For every schedule $\alpha =  v_1,\tau_1,v_2,\tau_2,v_3,\tau_3,\ldots$, let $F(\alpha)$ be the set of all $v \in S$ that occur only finitely often in $\alpha$ (\ie $v = v_i$ for finitely many $i \in \Nset_+$).
We put
\begin{equation*}
   \MP^C(\alpha)  =  \MP(\alpha) + \sum_{v \in C \cap F(\alpha)} c_v \,.
\end{equation*}
In other words, $\MP^C(\alpha)$ is the long-run average payoff per time unit when the agent commits to $\alpha$ and the set of compulsory nodes is~$C$. 
It may happen that $\MP^C(\alpha)$ is negative, which means that the agent is losing money in the long run and it should better not follow the schedule~$\alpha$.
Furthermore, we define
\begin{equation*}
   \Val^C  =  \sup_{\alpha}  \MP^C(\alpha)\,.
\end{equation*}
For a given $\varepsilon \geq 0$, we say that $\alpha$ is \emph{$\varepsilon$-optimal} if \mbox{$\Val^C - \MP^C(\alpha) \le\varepsilon$}.
In particular, $0$-optimal schedules are called \emph{optimal}.

\section{Deterministic Periodic Schedules}
\label{sec-periodic}

General schedules are not finitely representable and hence not apt for algorithmic purposes.
A workable subclass of deterministic schedules are \emph{periodic} schedules.
A schedule $\alpha$ is \emph{periodic} if there exists a finite \emph{generating cycle} $\beta = v_1,\tau_1,\ldots, v_n$ such that $v_1=v_n$ and $\alpha$ is the concatenation\footnote{For finite paths $\beta = v_1,\ldots,v_n$ and $\gamma = u_1,\ldots,u_m$ such that $v_n=u_1$, their concatenation $\beta \odot \gamma$ is the finite path $v_1,\ldots,v_n,u_2,\ldots,u_m$.} of infinitely many copies of $\beta$, i.e., $\alpha = \beta \odot \beta \odot \cdots$.
The \emph{length} of $\beta$ is defined as $n-1$.

Now we precisely formulate and prove the results~(1) and~(2) presented in Section~\ref{sec-intro}.
Since the proofs are not strictly needed for understanding the meaning and consequences of our theorems, they are given in Appendix~\ref{app-proofs}.

\begin{restatable}{theorem}{periodic}
   \label{thm-periodic}
   There exists an optimal periodic schedule such that the length of the generating cycle is at most exponential in $\size{\calS}$.    
\end{restatable}
   
Before stating our next theorem, we need to introduce some auxiliary notions. 
A service specification $\calS = (S,\tm,\{P_v : v \in S\},D)$ is \emph{simple} if $D= \emptyset$, $P_v(i) = i/|S|$ for all $i < k_v$, $c_v \leq 0$, and $d_v = 0$ (for every $v \in S$).
Observe that $\Val^C \leq 1$ for every simple $\calS$.
Furthermore, we say that the value of a simple $\calS$ is \emph{$\kappa$-separated} for a given $\kappa > 0$, if either $\Val^C = 1$ or $\Val^C \leq 1-\kappa$.
We have the following:

\begin{restatable}{theorem}{PSPACEhard}
\label{thm-PSPACE-hard}
Let $\kappa > 0$. The problem whether $\Val^C = 1$ for a given simple service specification $\calS$ with $\kappa$-separated value is \PSPACE-hard.   
\end{restatable}

Let us explain the consequences of Theorem~\ref{thm-PSPACE-hard}.
First, for every given $\varepsilon \geq 0$, the value $\Val^C$ of $\calS$ is not only hard to compute, but also hard to \emph{approximate} up to an additive error $\varepsilon$.
To see this, realize that the problem of Theorem~\ref{thm-PSPACE-hard}, i.e., the question whether $\Val^C = 1$, is equivalent to checking whether $\Val^C > 1 -\kappa/2$, because the value of $\calS$ is \mbox{$\kappa$-separated}.
Similarly, we obtain that the problem of constructing an $\varepsilon$-optimal periodic schedule is \PSPACE-hard.
Furthermore, for every fixed $\varepsilon \geq 0$, the length of the generating cycle $\beta$ of an $\varepsilon$-optimal periodic schedule $\alpha$ cannot be bounded by \emph{any} polynomial in $\size{\calS}$, unless $\mathsf{NP} = \mathsf{PSPACE}$ (if there was such a polynomial bound, then the problem of Theorem~\ref{thm-PSPACE-hard} would belong to \NP).
In fact, any \emph{subexponential} upper bound on the length of $\beta$ would bring unexpected consequences in complexity theory.
Hence, Theorem~\ref{thm-PSPACE-hard} is a strong evidence that the length of $\beta$ can be exponential in $\size{\calS}$, even for a simple $\calS$ with $\kappa$-separated value.

\section{RFM Strategies and Schedules}
  
In this section, we introduce \emph{randomized finite-memory (RFM) schedules} for the service agent.
Let $\calS = (S,\tm,\{P_v : v \in S\},D)$ be a service specification and $C$ a set of compulsory nodes.

\subsection{RFM Strategies}
\label{sec-RFM-definition}

Let $\Mem$ be a finite set of \emph{memory states}, and $\ag{S} = S \times M$ be the set of  \emph{augmented nodes}.
When denoting an augmented node by $\ag{v}$, we implicitly mean that $\ag{v} = (v,m)$ for some $m \in \Mem$.

A \emph{randomized finite-memory (RFM) strategy with memory $\Mem$} is a pair $(\sigma,\delta)$ where
\begin{itemize}
\item $\sigma\colon\ag{S} \to \calD(\ag{S})$ represents a randomized selection of the next node and the corresponding memory update;
\item $\delta\colon\ag{S} \times \ag{S} \to \Nset$ specifies a \emph{wait} for a given move. We require that $\delta(\ag{v},\ag{u}) > 0$ only if $(v,u) \in D$.  
\end{itemize}
Note that $(\ag{S},\sigma)$ can be seen as a Markov chain where $\ag{S}$ is the set of states and $\sigma$ is the transition function.
For every $\ag{v} \in \ag{S}$, let $\prob_{\ag{v}}$ be the probability measure in the standard probability space over all infinite sequences $\ag{v}_1,\ag{v}_2,\ag{v}_3,\ldots$ initiated in $\ag{v}$ (see, e.g., \cite{Norris:book}).
Furthermore, for every infinite sequence $\omega = \ag{v}_1,\ag{v}_2,\ag{v}_3,\ldots$, let $\alpha_{\omega} = v_1,\tau_1,v_2,\tau_2,\ldots$ be the associated deterministic schedule such that $\tau_i = \tm(v,v_{i+1}) + \delta(\ag{v}_i,\ag{v}_{i+1})$.
Then, we can interpret $\MP^C$ as a random variable over the infinite sequences in $(\ag{S},\sigma)$ by stipulating $\MP^C(\omega) = \MP^C(\alpha_{\omega})$.

\subsection{RFM Schedules}
\label{sec-RFM-schedules}

Let $(\sigma,\delta)$ be a RFM strategy with memory~$M$, and let $B$ be a bottom strongly connected component (BSCC) of $(\ag{S},\sigma)$.
Note that $B$ can be seen as an ergodic Markov chain,
and we use $\Inv_B(\ag{u})$ to denote the long-run average fraction of time units spent in $\ag{u}$.
By applying the standard results of ergodic theory (see, e.g., \cite{Norris:book}), we obtain that for every $\ag{v} \in B$, almost all infinite paths $\omega$ initiated in $\ag{v}$ satisfy
\begin{equation} \label{eq-MP}
	\MP^C(\alpha_{\omega})
		= \sum_{\ag{u} \in B} \Inv_B(\ag{u}) \cdot S(\ag{u})
			+ \sum_{v \in C \setminus S(B)} c_v,
\end{equation}
where we abbreviated
\begin{equation} \label{eq:inner}
	S(\ag{u}) = \sum_{t=1}^{\infty} \prob[\ag{u} \to_t u] \cdot P_u(t).
\end{equation}
Here, $\prob[\ag{u} \to_t u]$ is the probability that a path initiated in $\ag{u}$ visits an augmented vertex of the form $(u,m)$, where \mbox{$m\in \Mem$}, for the first time  after \emph{exactly} $t$~time units.
Furthermore, $S(B)$ is the set of all nodes $s$ such that $(s,m) \in B$ for some $m \in \Mem$.
Observe that~\eqref{eq-MP} is independent of~$\ag{v}$, and hence we can write just $\MP^C(B)$ to denote this value. 

Let $\{B_1,\ldots,B_n\}$ be the set of all BSCC of $(\ag{S},\sigma)$.
The \emph{value} of $(\sigma,\delta)$ is defined as 
\begin{equation}
   \Val(\sigma,\delta)  =  \max_{i \leq n} \{ \MP^C(B_i) \}.
\label{eq-Val}
\end{equation}
A BSCC $B$ is \emph{optimal for $(\sigma,\delta)$} if $\MP^C(B) = \Val(\sigma,\delta)$. 

Every optimal BSCC $B$ can be interpreted as a \emph{RFM schedule} for the service agent.
The agent starts in $v \in S(B)$ for some initial memory element $m$ such that $(v,m) \in B$, and proceeds by selecting the next moves and the associated waits according to $\sigma$ and $\delta$, respectively.
Although the concrete infinite path $\alpha$ obtained in this way is not a priori predictable, we have that $\MP^C(\alpha) = \Val(\sigma,\delta)$ for almost all~$\alpha$.
Formally, we define a \emph{RFM schedule} as a triple $(\sigma,\delta,B)$.

Note that every (deterministic) periodic schedule with a generating cycle $\beta$ can be represented as a RFM schedule with at most $|\beta|$ memory elements, where $|\beta|$ is the length of $\beta$.
The main advantage of RFM schedules is their \emph{compactness}, i.e., the capability of achieving a high mean payoff with a relatively low number of memory elements (see the example in Section~\ref{sec-intro}).

\section{Algorithms}
\label{sec-algo}

We describe our strategy synthesis algorithm for RFM schedules, and the algorithm for constructing a periodic schedule from a given RFM schedule by random sampling.

\subsection{Synthesizing RFM Schedules}

The strategy synthesis algorithm follows a machine learning approach.
At the beginning, we initialize $\sigma$ randomly.
Then, in an optimization loop, we compute the value $\Val(\sigma,\delta)$ and its gradient with respect to $\sigma$ and $\delta$
and we update them in the direction of the steepest ascent.
However, $\Val$ is not a differentiable function defined on an open set of real-valued parameters.
Indeed, the payoff functions is defined in discrete times $t$ and wait function $\delta$ is also constrained to a discrete set $\Nset$.
Therefore, we overcome non-differentiability by a different stochastic representation of waits $\delta$.

\subsubsection{Strategy Representation}
\label{subsec:waits-representation}

We ``split'' every prolongable edge $\ag{u}\to\ag{v}$ by inserting an auxiliary vertex $w$ (unique for each edge) and adding edges $\ag{u}\to w$, $w\to w$, $w\to\ag{v}$ with lengths $\tm(u,v)$, 1, 0, respectively.
In this new graph, the wait $\delta(\ag{u},\ag{v})$ is modeled by performing the self-loop on $w$ repeatedly $\delta(\ag{u},\ag{v})$-times.
Consequently, the waits are encoded in strategy $\sigma$ of the new graph and we use $\Val(\sigma)$ to denote the value produced by~$\sigma$.

Strategy $\sigma$ is a collection of probability distributions over outgoing edges at every vertex.
Hence, it ranges in a closed set of product of probability simplexes.
Therefore, we model every probability distribution of $\sigma$ by the \textsc{Softmax} function of a vector of unconstrained real-valued parameters.

\subsubsection{Evaluation}

By definition~\eqref{eq-Val}, we decompose $\sigma$ into strongly connected components using Tarjan's algorithm \cite{Tarjan:SCC-decomp-SICOMP}.
For each BSCC~$B$, we compute $\MP^C(B)$ from~\eqref{eq-MP} as follows.

\paragraph{Terms $\Inv_B$.}

First, we construct a system of $|B|+1$ linear equations with variables $(X_{\ag{u}})_{\ag{u}\in B}$,
where the first equation is $\sum_{\ag{u} \in B} X_{\ag{u}}=1$ and, for each $\ag{u}\in B$, we have the equation
\begin{equation*}
    X_{\ag{u}}=\sum_{\ag{v} \in B} X_{\ag{v}}\cdot\sigma(\ag{v})(\ag{u}).
\end{equation*}
Since $B$ is a single strongly connected component, this system has a unique solution $(x_{\ag{u}})_{\ag{u}\in B}$, corresponding to the invariant distribution, see, \eg~\cite{Norris:book}.
Next, for all $\ag{u},\ag{v}\in B$, we set $h_{\ag{u},\ag{v}}=x_{\ag{u}}\cdot\sigma(\ag{u})(\ag{v})$.
Observe that $h$ corresponds to the invariant distribution over the edges.
Then,
\begin{equation*}
	T=\sum_{\ag{u},\ag{v}\in B} h_{\ag{u},\ag{v}}\cdot\tm(u,v)
\end{equation*}
is the mean time of traversing an edge of $B$.
Finally, $\Inv_B(\ag{u})=x_{\ag{u}}/T$ is the relative amount of time spent on edges outgoing from $\ag{u}\in B$.
Note that the solutions $(x_{\ag{u}})_{\ag{u}\in B}$ depend smoothly on $\sigma$ and hence $\sigma\to \Inv_B$ is differentiable.
We implemented the computation in PyTorch library~\cite{PyTorch} where the gradients are calculated automatically.

\paragraph{Term $S(\ag{u})$.}

Computing the infinite sum from~\eqref{eq:inner} is somewhat harder.
Our solution is inspired by the technique designed in \cite{KKMR:Regstar-UAI} for evaluating strategies in patrolling games.
In our terminology, this evaluation algorithm inputs a RFM strategy $\sigma$, a vertex $v$ and a time threshold $k$, and it outputs, for each augmented node $\ag{w}$, the probability $\prob[\ag{w} \to_{\leq k} v]$ that $v$ is visited from $\ag{w}$ within $k$ time units.
Since this probability is computed as a sum $\sum_{t=1}^{k} \prob[\ag{w} \to_t v]$, we can use thisalgorithm to compute the probabilities $\prob[\ag{w} \to_t v]$ for each $t\in\{1,\dots,k\}$, and incorporate them into our evaluation function.  

\begin{algorithm}[t]
\small
\caption{Strategy optimization}
\label{alg:optim}
\begin{algorithmic}
\State ${\rm StrategyParams}\gets {\it Init}()$
\For{$i\in \{1,\ldots,{\rm Steps}\}$} 
	\State ${\rm Strategy} \gets {\it Softmax}({\rm StrategyParams})$
	\State ${\rm Value} \gets {\it Val}({\rm Strategy})$
	\State ${\rm Gradient} \gets {\it Gradient}({\rm Value})$
	\State ${\rm StrategyParams} += {\it Step}({\rm Gradient}, i)$
	\State \textbf{Save} ${\rm Value}, {\rm Strategy}$\vspace{0.5ex}
\EndFor
\Return ${\rm Strategy}$ with the largest ${\rm Value}$
\end{algorithmic}
\end{algorithm}
Thus, to compute $S(\ag{u})$, we run the procedure of~\cite{KKMR:Regstar-UAI} with $v=u$, $\ag{w}=\ag{u}$, $k=k_u-1$. If $c_u=d_u=0$ (\ie the payoff function $P_u$ is eventually zero), we are done.
If $c_u=0$ but $d_u\neq 0$ (\ie, $P_u$ is eventually constant), we have that
\begin{equation*}
	S(\ag{u})
		= \sum_{t=1}^{k_u-1} \prob[\ag{u} \to_t u] \cdot P_u(t)
		+ \sum_{t=k_u}^{\infty} \prob[\ag{u} \to_t u] \cdot P_u(t)
\end{equation*}
where $P_u(t)=P_u(k_u)$ for all $t\geq k_u$, and hence the latter sum is equal to $P_u(k_u)\cdot \sum_{t=k_u}^{\infty} \prob[\ag{u} \to_t u]$.
Since $\ag{u}$ lies in a \emph{bottom} strongly connected component, we have that the probability of revisiting $\ag{u}$ is $1$.
Hence, the probability of reaching $u$ from $\ag{u}$ is also $1$, and therefore
\begin{equation*}
	\sum_{t=k_u}^{\infty} \prob[\ag{u} \to_t u] =  1-\sum_{t=1}^{k_u-1} \prob[\ag{u} \to_t u]\,.
\end{equation*}
The most problematic case is when $c_u\neq 0$ and $P_u$ attains infinitely many different values.
We use the following trick.

\begin{proposition} \label{prop:deaffine}
For every $v \in S$, let $Q_v$ be a payoff function defined by $Q_v(i) = P_v(i) - i\cdot c_v$ for all $i \in \Nset$.
Then, for every schedule $\alpha = v_1,\tau_1,v_2,\tau_2,\dots$, we have that
\begin{equation} \label{eq:deaf}
   \MP^C(\alpha) = \MP[Q](\alpha) + \sum_{v \in C} c_v 
\end{equation}
where $\MP[Q](\alpha)$ denotes the mean payoff of $\alpha$ computed for the payoff functions $Q_v$.
\end{proposition}

Equality~\eqref{eq:deaf} follows easily by the definitions of $\MP^C$ and $\MP$.
Observe that for all $v\in S$ and $i\in\Nset$, we have that $Q_v(k_v{+}i) = d_v$, \ie the payoff functions $Q_v$ are \emph{eventually constant}.
Thus, Proposition~\ref{prop:deaffine} reduces the general case to the already considered case when $c_v=0$ for all $v\in S$.

We implemented a C++ extension of a PyTorch module containing these computations.
The gradient of $\prob[\ag{u} \to_t u]$ with respect to $\sigma$ is obtained by backpropagation.

\subsubsection{Optimization Loop}

The optimization loop (Algorithm~\ref{alg:optim}) is implemented using \textsc{PyTorch} framework~\cite{PyTorch} and its automatic differentiation with \textsc{Adam} optimizer~\cite{Adam}.
The strategy parameters are initialized at random by sampling from \textsc{LogUniform} distribution so that we impose no prior knowledge about $\sigma$.

\begin{algorithm}
	\small
	\caption{Sampling a periodic schedule from $\sigma$}
	\label{alg:determinize}
	\begin{algorithmic}
		\State ${\rm BestValue}\gets -\infty$
		\State $v[0]\gets\ag{v}$
		\For{$i\in\{0,1,\dots,s\}$}
			\State $(v[i+1],\delta)\gets{\it GetRandomSuccessor}(v[i])$
			\State $\tau[i]\gets\tm(v[i],v[i+1]) + \delta$
			\For{$j\in\{1,2,\dots,\min\{\ell,i+1\}\}$}
				\If{$v[i+1-j]=v[i+1]$}
					\State ${\rm CurValue}\gets{\it EvalSchedule}(v,\tau,i+1-j,i+1)$
					\If{${\rm CurValue}>{\rm BestValue}$}
						\State ${\rm BestValue}\gets{\rm CurValue}$
						\State ${\rm BestStrategy}\gets(i+1-j,i+1)$
					\EndIf
				\EndIf
			\EndFor
		\EndFor
		\State $(a,b)\gets {\rm BestStrategy}$
		\State \Return $[v[a],\tau[a],v[a+1],\tau[a+1],\dots,v[b-1],\tau[b-1],v[b]]$
	\end{algorithmic}
\end{algorithm}

\subsection{Computing Periodic Schedules}

We present Algorithm \ref{alg:determinize} for sampling a deterministic periodic schedule from a RFM schedule $\sigma$
where the waits are represented in the way explained in Section \ref{subsec:waits-representation}.
The algorithm inputs $\sigma$, a sample length $s$, an upper bound $\ell$ on the length of the generating cycle, and an initial augmented node $\ag{v}$.
${\it GetRandomSuccessor}(\ag{u})$ returns a successor of $\ag{u}$ and the corresponding wait $\delta$ chosen randomly according to~$\sigma$.
The function ${\it EvalSchedule}(v,\tau,a,b)$ computes the value of the periodic schedule with the generating cycle
\begin{equation*}
	v[a],\tau[a],v[a+1],\tau[a+1],\dots,v[b-1],\tau[b-1],v[b]
\end{equation*}
in the way specified in Algorithm~\ref{alg:eval_loop}.
The total running time of Algorithm~\ref{alg:determinize} is \mbox{$\calO(s\cdot(\log |\ag{S}|+\ell\cdot(|S|+\ell)))$}.
More details are given in Appedix~\ref{A:algorithms}.

\begin{algorithm}
	\small
	\caption{Evaluation of a periodic schedule}
	\label{alg:eval_loop}
	\begin{algorithmic}
	    \For{$u\in S$}
		    \State ${\rm FirstVisit}[u], {\rm LastVisit}[u]\gets none, none$
		\EndFor
		\State ${\rm Cost}, {\rm Length}, {\rm Penalty}\gets 0,0,0$
		\For{$i\in\{a,a+1,\dots,b\}$}
			\State $u\gets{\it DeAugmentify}(v[i])$
			\If{${\rm LastVisit}[u]=none$}
				\State ${\rm FirstVisit}[u]\gets {\rm Length}$
			\Else
			    \State ${\rm Cost}\gets {\rm Cost}+P_u({\rm Length}-{\rm LastVisit}[u])$
			\EndIf
			\State ${\rm LastVisit}[u]\gets {\rm Length}$
			\State ${\rm Length}\gets{\rm Length}+\tau[i]$
		\EndFor
		\For{$u\in S$}
		    \If{${\rm FirstVisit}[u]=none$}
		        \State ${\rm Penalty}\gets {\rm Penalty}+c_u$
		    \Else
			    \State ${\rm Cost}\gets {\rm Cost}+P_u({\rm Length}+{\rm FirstVisit}[u]$
					\State $\hspace*{13em}-{\rm LastVisit}[u])$
			\EndIf
		\EndFor
		\State \Return ${\rm Penalty}+{\rm Cost}/{\rm Length}$
	\end{algorithmic}
\end{algorithm}

\section{Experiments}

To demonstrate the functionality of our algorithms, we consider a set of parameterized service specifications of increasing size.
The code and experiments setup are available at
\smallskip

\centerline{\href{https://gitlab.fi.muni.cz/formela/2023-ijcai-periodic-maintenance}{gitlab.fi.muni.cz/formela/2023-ijcai-periodic-maintenance}.}

\subsection{Service specifications}

For each $k$, we construct a service specification $\calS_k$ consisting of one distinguished compulsory node (depot), $k$ nodes modeling machines with long maintenance time and high payoff, and $3k$ nodes representing machines with short maintenance time and lower payoff.
Time units are interpreted in minutes.

\paragraph{Payoffs.}

We aim to model an $8$-hour working shift.
We require that the agent returns to the depot before the end of a shift.
Here, the agent can receive only a (significantly) negative payoff for late arrival; earlier returns are not punished.
Hence, for the depot, we set $P(t)=0$ for $t\le 480$ and then $P$ decreases linearly with slope $-100$.
For each machine with long maintenance time, we set $P(t)=0$ for $t<6000$, $P(t)=6000$ for $6000\le t\le 7800$.
For $t>7800$, the payoff decreases by~$1$ for every time unit to stress their importance.
For machines with short maintenance time, we set $P(t)=0$ for all $t< 20$, $P(t)=1$ for $20\le t\le 40$, and $P(t)=0$ for $t>40$.
Note that the former machines are $20$ times more profiting (per time unit) than the latter ones when maintained on time.

\paragraph{Traversal Times.}

To avoid bias towards simple instances, we set traversal times randomly in the following way.
First, we position each machine into a $12\times 12$ grid randomly.
The depot node is always in the middle of the grid $(6,6)$.
For each pair of nodes $u$ and $v$ positioned at $(x_u,y_u)$ and $(x_v,y_v)$, we set the traversal time to
\begin{equation*}
   \tm(u,v) = 10\cdot(|x_u-x_v| + |y_u-y_v|)
\end{equation*}
that corresponds to the distance of the nodes in the grid, where each edge takes $10$~min.\ to traverse.

\subsection{Results}

\begin{table}[t]
	\newcolumntype{d}[1]{D{+}{\,$\pm$\,}{#1}}
	\centering\small
	\begin{tabular*}{\columnwidth}{@{\extracolsep{\fill}}rd{3,3}d{4,3}d{3,3}d{3,3}}
		\toprule
		 \multirow{2}{*}{$k$} & \multicolumn{2}{c}{Max Schedule Value} & \multicolumn{2}{c}{Runtimes [s]} \\ \cmidrule(lr){2-3}\cmidrule(lr){4-5}
			 & \multicolumn{1}{c}{Randomized} & \multicolumn{1}{c}{Periodic}
			 & \multicolumn{1}{c}{Alg.~\ref{alg:optim} (step)} 
			 & \multicolumn{1}{c}{Alg.~\ref{alg:determinize}} \\
		\midrule
		 2 & 0.24 + 0.12 &  1.42 + 0.51 &   0.01 + 0.0 &   11.44 + 1.56 \\
		 4 & 0.65 + 0.16 &  3.55 + 0.69 &   0.08 + 0.0 &    9.34 + 1.65 \\
		 6 & 1.17 + 0.04 &    6.0 + 0.0 &   0.28 + 0.0 &    8.62 + 1.41 \\
		 8 & 1.55 + 0.07 &  7.89 + 0.29 &   0.65 + 0.0 &    6.94 + 0.79 \\
		10 & 1.99 + 0.01 &   9.83 + 0.3 &  1.28 + 0.01 &    7.33 + 1.02 \\
		12 &  2.4 + 0.01 & 11.08 + 0.25 &  2.21 + 0.01 &    7.04 + 0.83 \\
		14 & 2.79 + 0.06 &  12.4 + 0.65 &  3.49 + 0.02 &    6.81 + 1.02 \\
		16 & 3.19 + 0.06 & 13.09 + 0.33 &  5.14 + 0.03 &    6.22 + 0.87 \\
		18 & 3.61 + 0.01 & 13.91 + 0.49 &   7.4 + 0.03 &    5.61 + 0.63 \\
		20 & 4.01 + 0.01 & 15.09 + 0.84 & 10.14 + 0.04 &    5.64 + 0.74 \\
		\bottomrule
	\end{tabular*}
	\caption{Maximal randomized (RFM) and periodic schedule value over 50 optimization steps of Alg.~\ref{alg:optim} on $\calS_k$ (mean over 12 restarts).}
	\label{tab-experiment}
\end{table}

\begin{figure}[t]
	\includegraphics[width=\columnwidth]{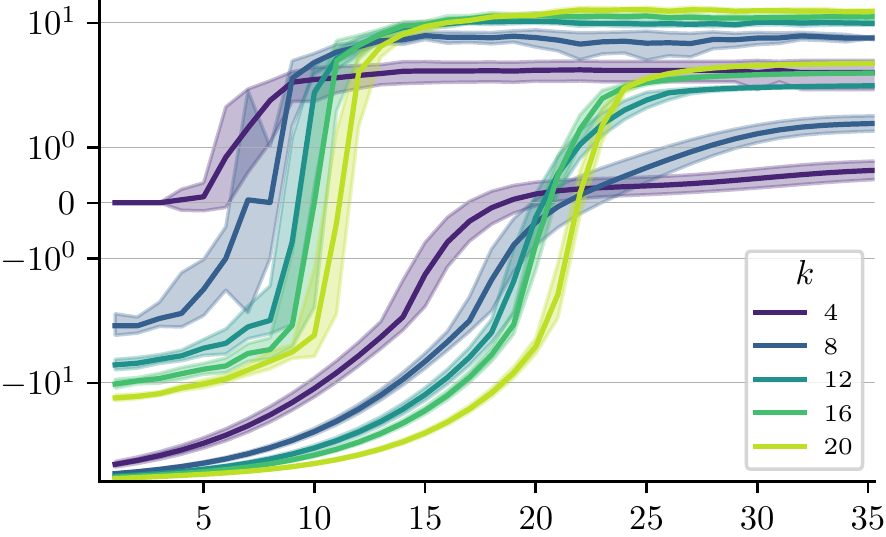}
	\caption{Optimization progress (mean and std over 12 restarts) for first 35 steps.
	Lower values correspond to the RFM schedule.
	Deterministic loops found by Alg.~\ref{alg:determinize} achieve significantly higher values.}
	\label{fig:convergence}
\end{figure}

We test our algorithm on $\calS_k$ with $k=2,4,6,\dots,20$.
For every $k$, we run Alg.~\ref{alg:optim} twelve times, each with $50$ optimization steps.
At every step, we take the current RFM schedule, and we feed it into Alg.~\ref{alg:determinize} to obtain a deterministic periodic schedule.\footnote{We used $s=10^5$, $\ell=300$ and the depot as an initial vertex.}
Therefore, for every run, we obtain the maximal value of RFM and periodic schedules achieved during the optimization.
The statistics are reported in Tab.~\ref{tab-experiment} together with average runtimes.
The optimization progress is also plotted in Fig.~\ref{fig:convergence} for selected $k$.
If every machine is visited ideally, the agent earns $k\cdot 1+3k\cdot\tfrac{1}{20}=k\cdot 1.15$.
The topology of the graph may not allow the agent to achieve such a payoff, but it serves as a reasonable upper bound.

We observe stable and convergent behaviour of both optimization (Alg.~\ref{alg:optim}) and determinization (Alg.~\ref{alg:determinize}) results.
We see that RFM optimization stabilizes around step 30 on relatively small values.
The periodic schedules achieve significantly higher scores consistently, and their quality improves together with the quality of the initial RFM.
Interestingly, the maxima are attained much earlier (around step 15).
This shows that a slight initial improvement of the RFM schedule leads to a significant boost of a sampled periodic schedule.

A more detailed analysis is in Appendix~\ref{A:experiments}.
We mention that the periodic schedules are typically 100-200 nodes long with traverse time exceeding 12 working shifts.
All RFMs were optimized with memory size 1, which explains their lower values.
Experiments with larger memory are also in Appendix~\ref{A:experiments}.
In conclusion, to find periodic schedules as RFM, the memory has to be large (to encode, say, 200 long cycle on $4k+1$ nodes), which is expensive for larger graphs.
Smaller memories above 1 do not help significantly.
Therefore optimizing memoryless RFM combined with determinization gives satisfactorily high values with very low runtimes.

\section{Conclusion}
\label{sec-concl}

We introduced the infinite horizon recurrent routing problem (IHRRP) and presented two solution concepts: randomized finite memory (RFM) and periodic schedules.
In the theoretical part, we proved that even the problem of bounding the value of an optimal periodic schedule is \PSPACE-hard.
Hence, we proposed an algorithmic solution based on randomized algorithms.
We apply differentiable programming with optimization techniques to synthesize promising RFM schedules that, by random sampling procedure, yield strong periodic schedules.
We demonstrated on reasonably-sized examples that our approach stably produces long schedules that achieve a high mean payoff.

\section*{Acknowledgments}

This work has been supported by the Czech Science Foundation, Grant No.~21-24711S.

\appendix

\begin{table}[t]
	\newcolumntype{d}[1]{D{+}{\,$\pm$\,}{#1}}
	\centering\small
	\begin{tabular}{rd{3,2}d{4,3}rd{3,2}d{4,3}}
		\toprule
		 \multirow{2}{*}{$k$}
		 	 & \multicolumn{2}{c}{Schedule}
			 	& \multirow{2}{*}{$k$}
				 & \multicolumn{2}{c}{Schedule}
					 \\ \cmidrule(lr){2-3} \cmidrule(lr){5-6}
			 & \multicolumn{1}{c}{Length} & \multicolumn{1}{c}{Time}
				 & & \multicolumn{1}{c}{Length} & \multicolumn{1}{c}{Time} \\
		\cmidrule(r){1-3} \cmidrule(l){4-6}
		 2 & 142 + 59 &   6000 + 2 & 12 & 141 + 30 & 6030 + 112 \\
		 4 & 172 + 38 &  6009 + 74 & 14 & 140 + 20 & 6109 + 182 \\
		 6 & 162 + 32 &   6000 + 0 & 16 & 150 + 26 & 6112 + 159 \\
		 8 & 168 + 42 &  6006 + 38 & 18 & 150 + 33 & 6184 + 152 \\
		10 & 138 + 41 & 6100 + 192 & 20 & 157 + 28 & 6363 + 363 \\
		\bottomrule
	\end{tabular}
	\vskip-1ex
	\caption{Analysis of deterministic periodic schedules found by Alg.~2 achieving maximal values.
	Length denotes the number of schedule's vertices and Time denotes total traverse time.
	Schedules are consistently more than 12 working shifts long.
	}
	\label{tab-schedules}
\end{table}

\begin{figure}[b]
	\centering
	\includegraphics[width=\columnwidth]{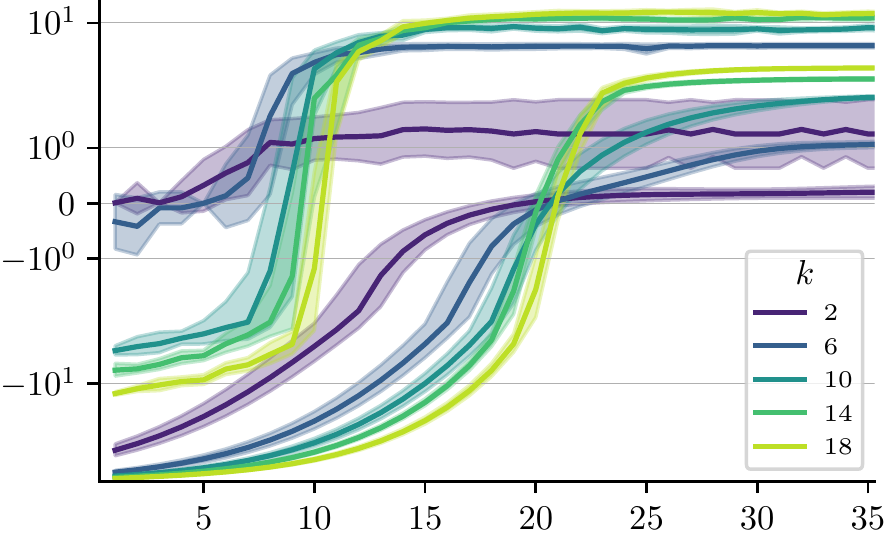}
	\vskip-1ex
	\caption{Optimization progress (mean and std over 12 restarts) for first 35 steps.
	Lower values correspond to the RFM schedule.}
	\label{fig:convergence_remaining}
\end{figure}

\section{Experimental Details} \label{A:experiments}

\paragraph{Additional Data}

Figure~\ref{fig:convergence_remaining} shows optimization progress (Alg.~1) and determinization (Alg.~2) for additional graph parameters $k$.
We see very similar behaviour.

\paragraph{Periodic Schedule Analysis}

We analysed periodic schedules found by Alg.~2.
For every schedule, we count the number of visited vertices (Length) and its total traverse time (Time).
In Tab.~\ref{tab-schedules}, for each graph parameter $k$, we report the means of all schedules that attained maximal value.
Note, that in each of the 12 restarts, the maxima is usually attained more than once.

\paragraph{Memory Size Analysis}

\begin{figure}[t]
	\centering
	\includegraphics[width=\columnwidth]{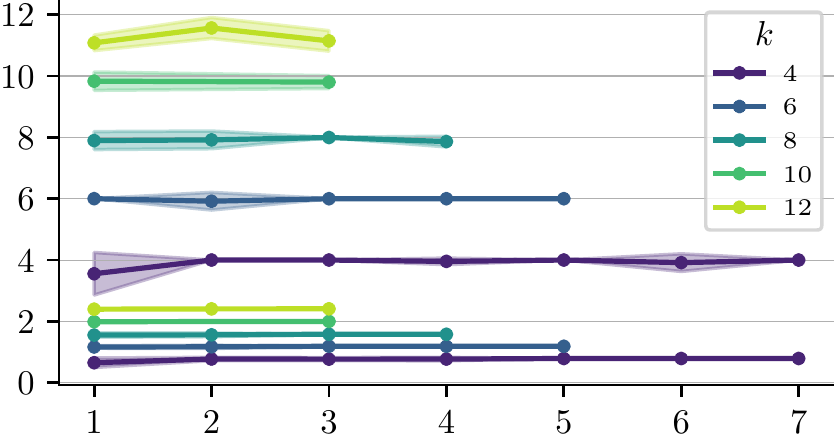}
	\vskip-1ex
	\caption{Best RFM and periodic schedule values achieved with increasing RFM memory size 1-7 and various graph parameters~$k$ (mean and std over 12 restarts).
	Lower values correspond to RFM schedules.
	Larger memory does not have impact on performance.}
	\label{fig:memory}
\end{figure}

\begin{figure}[b]
	\centering
	\includegraphics[width=\columnwidth]{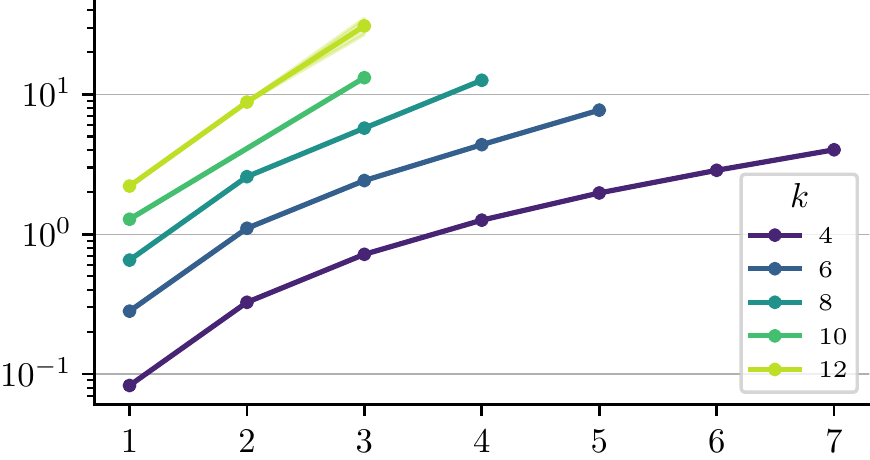}
	\vskip-1ex
	\caption{Runtime (in seconds) of a single step of optimization Alg.~1 for increasing RFM memory size 1-7 and various graph parameters~$k$.
	}
	\label{fig:memory_times}
\end{figure}

We tested various memory sizes on the same grid graphs instances.
For each $k$ and memory size $1,\ldots,7$, we run 50 steps of Alg.~1 and Alg.~2, all with 12 restarts.
Figure~\ref{fig:memory} summarizes the results.
Some runs did not finished due to timeouts.
We conclude that larger memory does not lead to higher values.
In contrast, the runtime increase is significant as reported in Fig.~\ref{fig:memory_times}.

\section{Algorithm Details} \label{A:algorithms}

\paragraph{Function ${\it GetRandomSuccessor}(\ag{u})$.}

It returns a successor of $\ag{u}$ and the corresponding wait $\delta$ chosen randomly according to $\sigma$.
This is implemented by picking a random number $r\in[0;1)$ and finding the least $i$ satisfying 
\begin{equation*}
  \sum_{j=1}^{i}\,\sigma(\ag{u})(v_j)>r,
\end{equation*}
where $[v_1,v_2,\dots,v_q]$ is a precomputed list of all successors of $\ag{u}$. For a large $q$, instead of finding the~$i$ in a simple loop in time $\Theta(q)$, we precompute the prefix sums $\sum_{j=1}^{z}\,\sigma(\ag{u})(v_j)$ for each $z=0,1,\dots,q$ and
then find the $i$ by binary search, reducing the time to $\Theta(\log q)$ (note that $q\leq |\ag{S}|$).

If the chosen successor is a standard node (\ie, it is not an auxiliary vertex $w_{\ag{u}\to\ag{v}}$), we return $\delta=0$.
If the successor is $w_{\ag{u}\to\ag{v}}$, we could simply iterate the procedure until the chosen successor is $\ag{v}$ and set $\delta$ to the number of iterations.
However, there is a quicker way, resulting in the same distribution of $\delta$:
Let $p$ denote the probability of the loop on $w_{\ag{u}\to\ag{v}}$ in $\sigma$.
We pick a random number $r\in(0;1]$ and set 
\begin{equation*}
	\delta=\lfloor\log_{p}r\rfloor=\lfloor\ln r/\ln p\rfloor.
\end{equation*}

\paragraph{Time Complexity of Computing Periodic Schedules}

Clearly, ${\it EvalSchedule}$ runs in time
$\Theta(|S|+(b-a))$, which is $\calO(|S|+\ell)$.
Since ${\it GetRandomSuccessor}$ runs in $\calO(\log |\ag{S}|)$,
we get that the time complexity of the entire Algorithm 2
is \mbox{$\calO(s\cdot(\log |\ag{S}|+\ell\cdot(|S|+\ell)))$}.

One could further get rid of the $s\ell^2$ term by utilizing the fact that, having computed
the visit times in \textsc{EvalSchedule} for some $(a,b)$, then in the next call for $(a',b)$, where~$a'<a$,
it is enough to recompute the visits which go through the newly added cycle from~$v[a']$ to~$v[a]$.

\section{Proofs}
\label{app-proofs}
\periodic*
 
\begin{proof}
Let  $C \subseteq S$ be a set of compulsory nodes. Recall that $c_v = 0$ for all $v \in S {\smallsetminus} C$.
For every $v \in S$, we define a new payoff function $Q_v$ as follows: 
\begin{equation}
  Q_v(i) = P_v(i) - i\cdot c_v
	\quad \text{for all $i \in \Nset$}. 
\end{equation}

Observe that $Q_v(k_v{+}i) = d_v$ for all $v \in S$ and $i \in \Nset$.
Furthermore, for every schedule $\alpha = v_1,\tau_1,v_2,\tau_2,\dots$,
\begin{equation}
   \MP^C(\alpha) = \MP[Q](\alpha) + \sum\nolimits_{v \in C} c_v,
\end{equation}
where $\MP[Q](\alpha)$ denotes the mean payoff of $\alpha$ computed for the payoff functions $Q_v$.
Hence, maximizing $\MP^C(\alpha)$ is equivalent to maximizing $\MP[Q](\alpha)$. 

Recall that the $Q$-payoff received at~$v$ is the same whenever $v$ is revisited after $k_v$ or more time units.
Let $T$ be the (finite) set of all $f\colon S \rightarrow \Nset_+$ such that $f(v) \leq k_v$ for all $v \in S$.
Consider a directed graph $G$ where $S \times T$ is the set of vertices, and $(v,f) \to (u,g)$ is an edge iff there exists $w \geq 0$ such that 
\begin{equation*}
   g(s) = \begin{cases}
            \min\{k_s,f(s)+\tm(v,u)+w\} & \text{if } v \neq s,\\ 
            \min\{k_s,\tm(v,u)+w\}      & \text{if } v=s.
          \end{cases}
\end{equation*}
We also require that $w > 0$ only if $(v,u) \in D$.
To every $(v,f)$ we assign the payoff $Q_v(f(v))$.
Clearly, every schedule $\alpha$ determines a unique infinite path in $G$, and vice versa.
Furthermore, $\MP[Q](\alpha)$ is equal to the mean payoff of the corresponding path in $G$.
By applying standard results for mean payoff games (see, e.g., \cite{EM:mean-payoff-JGT}), we obtain that the optimal path in $G$ exists and has a form of a `lasso', \ie, a finite path followed by a \emph{simple} cycle (where no vertex is visited more than once) that is iterated infinitely often.
Since the agent is free to choose the initial node in the periodic schedule, we may take the simple cycle to be the generating cycle of the periodic schedule (starting at any of its nodes). This may lead to obtaining a different payoff when a node is visited for the first time, but this will happen only finitely many times, so the mean payoff will stay the same.
Since the length of the cycle is bounded by $|S {\times} T|$, we obtain that the length of the corresponding optimal periodic schedule in $\calS$ is at most exponential in~$\size{\calS}$.
\end{proof}

\PSPACEhard*
\begin{proof}
The result is obtained by reduction from the \emph{cyclic routing UAV problem (cr-UAV)} which is \PSPACE-complete \cite{HO:UAV-problem-PSPACE}.
An instance of cr-UAV is a tuple $(V,\tm,\delta)$ where $V \neq \emptyset$ is a set of vertices, $t\colon V \times V \to \Nset_+$ are traversal times, and $\delta\colon V \to \Nset_+$ assigns to each $v \in V$ the \emph{deadline} for revisiting~$v$.
All traversal times and deadlines are bounded by $f(|V|)$ where $f\colon \Nset \to \Nset$ is a fixed quadratic function independent of the instance.
The question is whether there exists an \emph{eligible} sequence $\gamma = v_1,v_2,\ldots$ such that every $v \in V$ is visited infinitely often along $\gamma$, and the time between two consecutive visits to $v$ is always at most $\delta(v)$.
Let $\kappa > 0$, and let $(V,\tm,\delta)$ be an instance of cr-UAV. Furthermore, let $\varrho = 1+ |V| \cdot \prod_{v \in V} \delta(v)$.
Observe that the value of $\varrho$ is at most \emph{singly exponential} in $|V|$, and hence the length of the binary encoding of $\varrho$ is \emph{polynomial} in $|V|$. 

Let $\calS = (S,\tm,\{P_v : v \in S\},D)$ where $S = V$, $\tm = t$, $k_v = \delta(v)+1$, $P_v(i) = i/|S|$ for all $i < k_v$, $d_v = 0$, $c_v = - (\kappa + 2 \cdot k_{\max}\cdot \varrho)$, and  $D = \emptyset$. 
Observe that $\calS$ is simple and $\size{\calS}$ is polynomial in $|V|$. All nodes of $S$ are declared compulsory, i.e., $C = S$. We prove the following:
\begin{itemize}
\item[(a)] If there is an eligible sequence $\gamma = v_1,v_2,\ldots$ in the considered instance of cr-UAV, then $\Val^C =1$ in~$\calS$.
\item[(b)] If there is no eligible sequence $\gamma = v_1,v_2,\ldots$ in the considered instance of cr-UAV, then $\Val^C \leq 1-\kappa$ in~$\calS$.
\end{itemize}
Claim~(a) is immediate, because $\gamma$ determines a (unique) schedule $\alpha$ satisfying $\MP^C(\alpha)=1$. 

To prove Claim~(b), assume that there is no eligible infinite sequence of vertices in $(V,\tm,\delta)$, and consider an arbitrary schedule $\alpha =  v_1,\tau_1,v_2,\tau_2,\ldots$ in $\calS$. We show that $\MP^C(\alpha) \leq  1-\kappa$. 

If some  $u \in S$ is visited only finitely often along $\alpha$, we have that $\MP^C(\alpha) \leq 1 + c_u < 1 - \kappa$. 
Now assume that all nodes are visited infinitely often along~$\alpha$.

For every $i \in \Nset_+$, let $\eta_i\colon S \rightarrow \Nset_+$ such that $\eta_i(v)$ is the total time since the previous visit to $v$ before visiting $v_i$, i.e., $\eta_i(v) = \sum_{j=k}^{i-1} \tau_j$ where $k$ is the largest index $m$ such that $m<i$ and $v_m = v$ (if there is no such $m$, we put $k=1$). We say that $\eta_i$ is \emph{late} if $\eta_i(u) \geq k_u$ for some $u \in S$. 

Recall that $\varrho = 1+ |S| \cdot \prod_{v \in S} (k_v-1)$ and observe that the number of
different $\eta$'s that are not late is bounded above by $\varrho$.
We prove that for every $i \geq 1$, there is $j < \varrho$ such that $\eta_{i+j}$ is late.
Assume the converse, i.e., there is $i \geq 1$ such that $\eta_{i+j}$ is \emph{not} late for every $j < \varrho$.
Then, by the pigeonhole principle, there exist indexes $\ell,\ell'$ such that $i \leq \ell < \ell' < i+\varrho$, $v_{\ell} = v_{\ell'}$, and $\eta_{\ell} = \eta_{\ell'}$.
Consider the cycle $v_\ell,\ldots,v_{\ell'}$.
Clearly, every node is visited along this cycle and the time elapsed between two consecutive visits to the same node $u$ along this cycle is at most $\delta(u)$.
Hence, by iterating the cycle, we obtain an eligible infinite path in  $(V,\tm,\delta)$, which is a contradiction.

According to the above paragraph, for every $i \geq 1$, there is $j < \varrho$ such that $\eta_{i+j}$ is late.
This means that some vertex receives a large negative payoff in the next visit.
We show that this negative payoff is large enough to achieve $\MP^C(\alpha) \leq  1-\kappa$.
To see this, we split $\alpha$ into infinitely many finite paths of the form $\alpha = \lambda_1 \odot \theta_1 \odot \lambda_2 \odot \theta_2 \ \cdots$ where 
\begin{itemize}
\item $\lambda_{i+1}$ is the shortest path executed in $\alpha$ after $\lambda_1 \odot \cdots \odot \theta_i$ ending in a vertex $v_j$ such that $\eta_j$ is late;
\item $\theta_{i+1}$ is the shortest path executed in $\alpha$ after $\lambda_1 \odot \cdots \odot \lambda_{i+1}$ ending in a vertex $u$ such that $\eta_j(u) \geq k_u$, where $v_j$ is the last vertex of $\lambda_{i+1}$ (such $u$ exists, since we assume that each node is visited infinitely often along~$\alpha$).
\end{itemize}
Observe that the length of every $\lambda_i$ is at most $\varrho$.
For every $j \geq 1$, let $N_j$ be the length of $\lambda_1 \odot \theta_1 \odot \cdots \odot \lambda_{j} \odot \theta_{j}$.
Clearly,
\begin{equation}
   \MP^C(\alpha) ~
		 \le ~ \liminf_{j \to \infty} \frac{\sum_{i=1}^{N_j} P_{v_i}(t_i)}{\sum_{i=1}^{N_j} \tau_i} \,,
\label{eq-liminf}
\end{equation}
where $t_i$ denotes the total time elapsed since the previous visit to $v_i$
(\cf Subsection \ref{sec-schedules}).
For every $j \geq 1$, we have that
\begin{equation*}
	\sum_{i=1}^{N_j} P_{v_i}(t_i)  ~\le~  \frac{k_{\max}}{|S|}\cdot \varrho \cdot j + \left(\frac{k_{\max}}{|S|} + c_{\max}\right) \cdot j
\end{equation*}
and $\sum_{i=1}^{N_j} \tau_i  \ge  j$.
Hence, the fraction in \eqref{eq-liminf} is bounded by $1-\kappa$ for all~$j$.
\end{proof}

{\fontsize{10pt}{10.67pt}\selectfont

}

%%%%%%%%%%%%%%%%%%%%%%%%%%%%%%%%%%%%%%%%%%%%%%%%%%%%%%%%%%%%%%%%%%%%%%%%

\end{document}